\documentclass[final,english]{bullsrsl}[2022/06/15]

\usepackage[latin1]{inputenc}
\usepackage[T1]{fontenc}

\usepackage{natbib} 
\usepackage{graphicx}

\newcommand\dsct{$\delta$\,Sct}
\newcommand\teff{$T_{\rm eff}$}
\newcommand\nua{$\nu_{1}$}
\newcommand\nub{$\nu_{2}$}

\begin{document}
\title{Amplitude Modulation in a $\delta$ Sct star HD\,118660 }

\author[affil={1,2},corresponding]{Mrinmoy}{Sarkar}
\author[affil={1}]{Santosh}{Joshi}
\author[affil={3}]{Peter}{De Cat}
\affiliation[1]{Aryabhatta Research Institute of Observational Sciences (ARIES), India }
\affiliation[2]{M.J.P Rohilkhand University, Bareilly, Uttar Pradesh-243006, India }
\affiliation[3]{Royal Observatory of Belgium, Belgium}
\correspondance{mrinmoysarkar@aries.res.in}
\date{31 May 2023}
\maketitle

\begin{abstract}
  
In this paper, we report the detection of amplitude modulation in a $\delta$\,Scuti star HD\,118660.  We found that the $p$-mode frequency at $24.3837~d^{-1}$ varies periodically in amplitude with frequency $0.0558\pm0.00147d^{-1}$.  
However, all other modes are stable in both amplitude and phase which is clear evidence of non-conservation of visible pulsation mode energy.  
We constructed a two-frequency model by superimposing two sinuso\"ids with frequencies $\nu_{1} = 24.3837 d^{-1}$ and $\nu_{2} = 24.4420d^{-1}$ and corresponding phases $\phi_{1} = 0.5211$ rad and $\phi_{2} = 0.9481$ rad to mimic the observed variations of amplitude and phase with time.
The plausible explanation of the amplitude modulation in HD\,118660 is due to beating of two unresolved closed frequencies $\nu_{1}$ and $\nu_{2}$. 

\end{abstract}


\section{Introduction}
HD\,118660 is one of the targets of the ``Nainital-Cape'' (N-C) survey project, one of the longest ground-based surveys to search for and study the photometric variability in two classes of chemically peculiar (CP) stars, namely the Ap and Am stars. Apart from having chemical peculiarity, this star also shows multi-periodic behavior in the frequency spectrum, which makes it a good candidate for investigation. 
For this survey, a well-defined sample of more than 350 targets was surveyed, leading to the detection of $\delta$\, Scuti-type pulsational variability in eight Am stars (including HD\,188660) and rapid pulsation in one Ap star \citep{2000ashoka,2001martinez,joshi2003,2006joshi,joshi2009,joshi2010,joshi2012,joshi2016,joshi2017, 2022MNRAS.510.5854J}.  

The $\delta$\,Scuti (\dsct) stars are a class of pulsating variables situated in the lower part of the classical instability strip in the Hertzsprung-Russell diagram. 
These stars pulsate with various radial and non-radial $p$-mode \citep{2011A&A...534A.125U}. 
The spectral type of these stars ranges from A2 to F2. The effective temperature (\teff) of these main-sequence stars lies between 7030\,K and 9040\,K. 
The pulsations of \dsct\ stars are excited by a heat engine driving mechanism caused by an increased opacity in their surface layers \citep{2017ampm.book.....B}. They have pulsation periods of order a few hours \citep{2015JApA...36...33J}. 

Space-based observations of \dsct\ stars in the last decade have revolutionized our knowledge by providing ultra-precise photometric data with a duty cycle of close to 100\%. Such data allows for studying the internal stellar structure and evolution of low- and high-mass stars in more detail. 

\begin{figure}[ht]
\centering
\includegraphics[width=16cm]{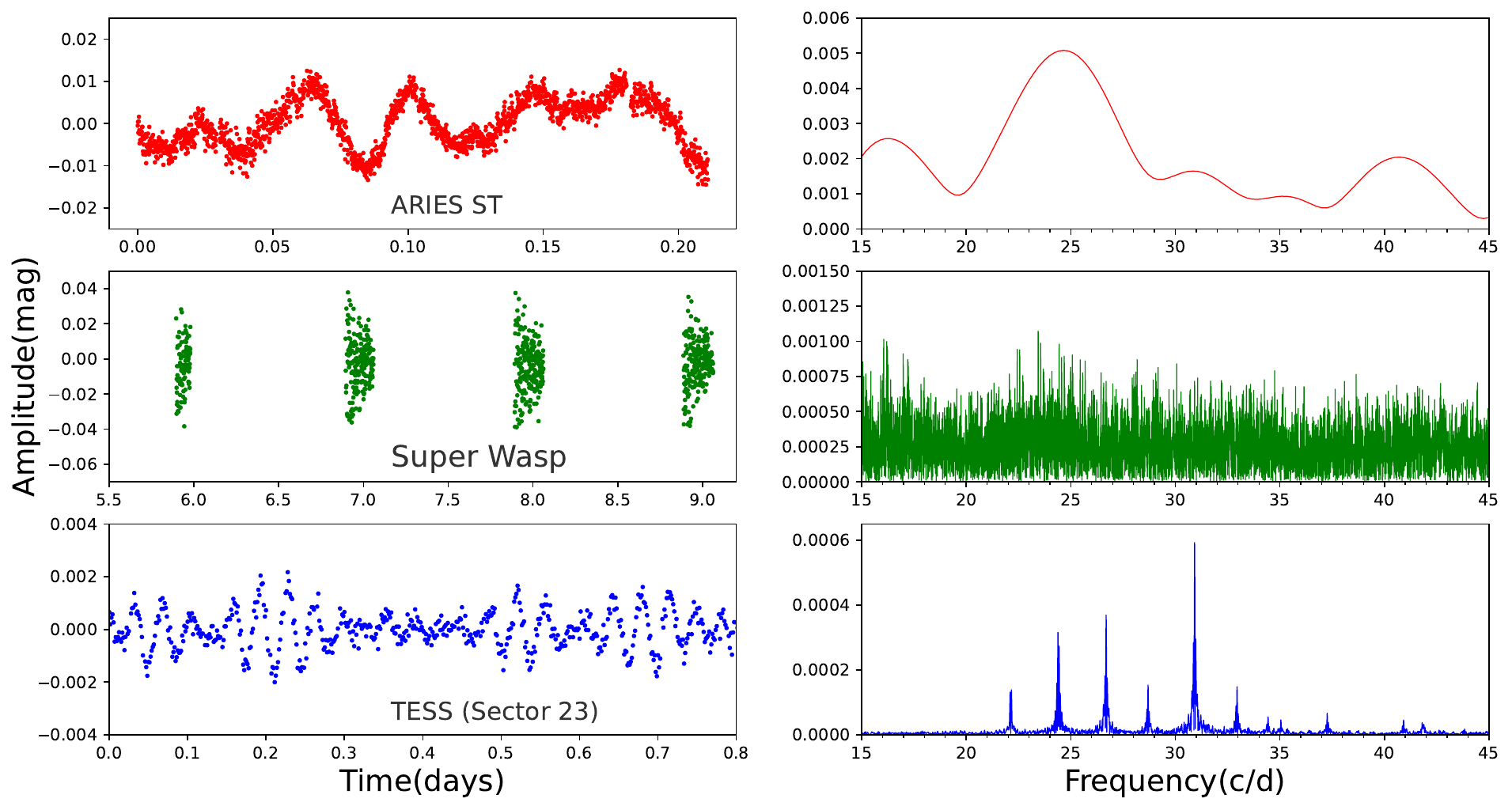}
\bigskip
\begin{minipage}{12cm}
\caption{
The light curves of HD\,118660 (left panels) and corresponding frequency spectra (right panels) were obtained with the 1.04-m Sampurnanand telescope of ARIES in the Johnson B band (top row), SuperWASP (middle row), and TESS (bottom row).
}
\label{fig:lc}
\end{minipage}
\end{figure}

HD\,118660 was classified as a multi-periodic \dsct\ pulsator by \cite{2006joshi}.
The top panels of Fig.\,\ref{fig:lc} show the discovery light curve obtained with the 1.04-m Sampurnanand telescope of ARIES on February 24, 2005, under the N-C survey project (left) and the corresponding periodogram (right). 
Later, this star was also observed with the ground-based telescope SuperWASP on January 30, 2013 \citep{2006pollacco}. 
The SuperWASP light curve and corresponding amplitude spectrum are shown in the middle left and right panels of Fig. \ref{fig:lc}, respectively. 
Recently, HD\,118660 was monitored by the space mission Transiting Exoplanet Survey Satellite (TESS; \citealt{2015JATIS...1a4003R}) in sectors 23 and 50.  
The TESS light curve of HD\,118660 observed in sector 23 and the corresponding amplitude spectrum are shown in the bottom panels of Fig. \ref{fig:lc}. 
This figure clearly demonstrates the difference in the photometric precision of data obtained from ground-based observing sites and those obtained from space. 
In addition to the precision of the observation, the length of the time base of the data is equally important for a precise determination of asteroseismic parameters such as the frequencies of the observed pulsation modes.  
In this manuscript, we present the analysis of the time-series data of HD\,118660 obtained with TESS.

\section{Observations and Data Analysis}\label{obs}

TESS is an all-sky survey launched by NASA on April 18, 2018, with the detection of exoplanets using the transit method as the primary goal, but it also provides data for pulsating variables ideal for asteroseismic studies. 
It observes sectors with a size of 24$^{\circ}$\,$\times$\,96$^{\circ}$. So far, 83 sectors are defined up to October 1, 2024. 
Each sector is observed for 27 days with a cadence of 30\,min (long cadence) and 2\,min (short cadence). 
HD\,118660 has been observed in sectors 23 (starting on March 18, 2020) and 50 (starting on March 26, 2022) in short cadence mode. 
Two types of TESS photometry are available: the simple aperture photometry (SAP) flux and the pre-search data conditioning SAP (PDCSAP) flux.
The PDCSAP is corrected for long-term trends mainly attributed to  instrumental effects. 
In our study, we have used the 2-min cadence PDCSAP light curves downloaded from the Barbara A. Mikulski Archive for Space Telescopes (MAST) using the python package \emph{lightkurve} \citep{2018ascl.soft12013L}. 

The PDCSAP light curves of sectors 23 and 50 were subjected to a frequency analysis to detect and identify the dominant frequencies using the $Lomb-Scargle$ method \citep{1982ApJ...263..835S} implemented in the \texttt{PERIOD04} package developed by \cite{2014ascl.soft07009L}.  
We selected frequencies with amplitudes having a signal-to-noise ratio (SNR) above 4. 


\section{Amplitude Modulation}
\label{indi}

It is found that the majority of \dsct\ stars exhibit amplitude modulation where the amplitude of at least one pulsation mode changes over time scales of the order of years to decades \citep{2016MNRAS.460.1970B}.
The amplitude modulation in \dsct\ pulsating variables could be due to either beating of close-frequencies pulsation modes \citep{2002A&A...385..537B} or mode-coupling between different combinations of frequencies \citep{2000ASPC..210....3B}. 
There is a large number of pulsating stars observed by TESS in more than one sector for which amplitude modulation is worth to be investigated.
HD\,118660 is one of them.
\begin{figure}[ht]
\centering
\includegraphics[width=14cm]{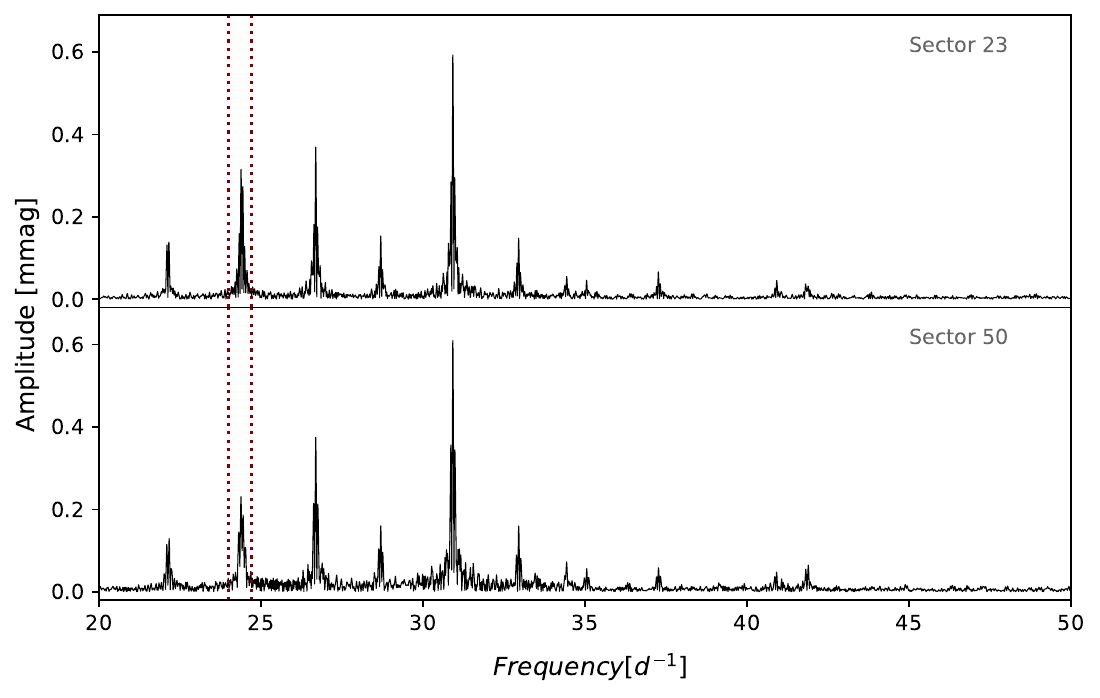}
\bigskip
\begin{minipage}{12cm}
\caption{Comparison of the amplitude spectra of HD\,118660 observed in TESS sector 23 (upper panel) and 50 (lower panel). The frequency peak in between the \textit{maroon} dotted lines, corresponding to \nua\,=\,24.3837\,d$^{-1}$, shows amplitude modulation.}
\label{fig:compare_observe}
\end{minipage}
\end{figure}

\begin{figure}[ht]
\centering
\includegraphics[width=14cm]{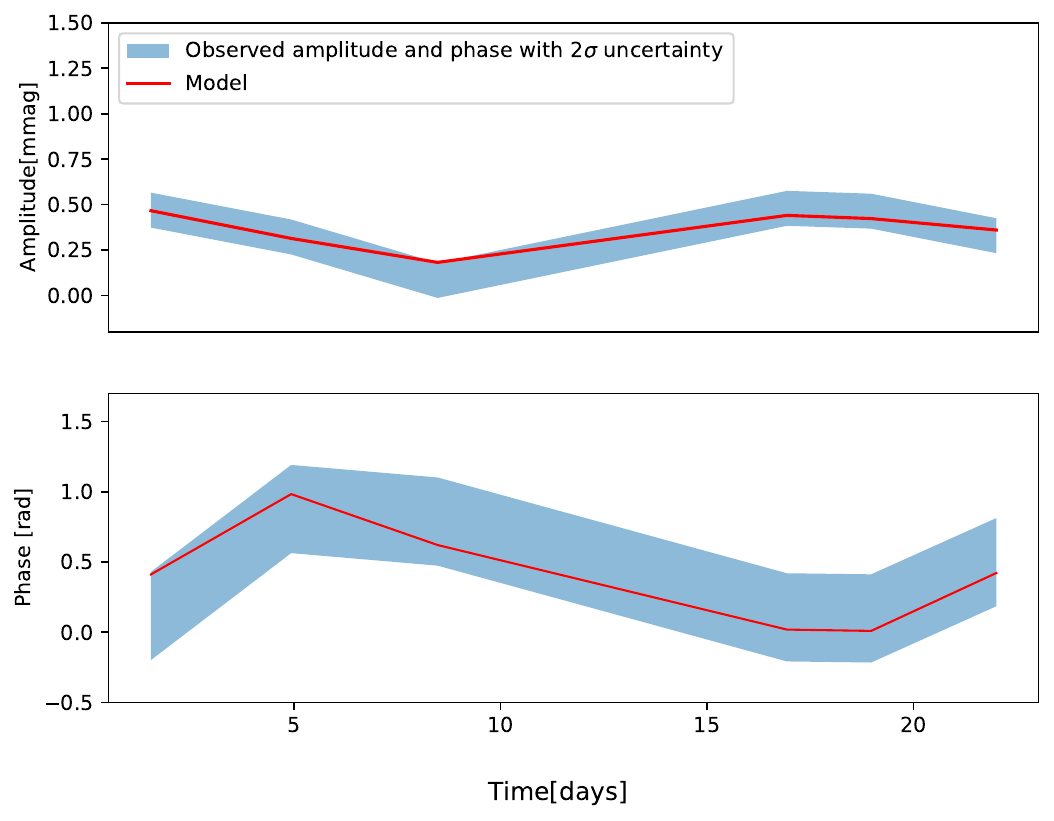}
\bigskip
\begin{minipage}{12cm}
\caption{
The blue region represents the calculated amplitude (\textit{upper panel}) and phase (\textit{lower panel}) at  frequency \nua\,=\,24.3837$\pm$\,0.0005\,d$^{-1}$ from TESS sector 23 observation with $2\sigma$ spread from the model (\textit{red line}).
}
\label{fig:amp_model}
\end{minipage}
\end{figure}

Amplitude modulation can be visualised by comparing the amplitude of light variations in a given time interval. 
The comparison of the amplitude spectra for sectors 23 and 50 are shown in Fig.\,\ref{fig:compare_observe} where the amplitude of the $p$-mode frequency \nua\,=\,24.3837$\pm 0.0005 \,d^{-1}$ decreased from 0.3169$\pm 0.0075$ to 0.2309$\pm 0.0091$\,mmag over a period of 735\,days. To inspect the amplitude and phase variation against time, we divided the entire data set into bins of one day each and calculated the amplitude spectrum for each bin individually. The change of amplitude and phase with time for \nua\ are shown in Fig.\,\ref{fig:amp_model}.
Interestingly, the variations have the same period but are anti-phase. This could be attributed to the presence of another pulsation frequency close to \nua\ \citep{2016MNRAS.460.1970B}. 

\begin{figure}[ht]
\centering
\includegraphics[width=\textwidth]{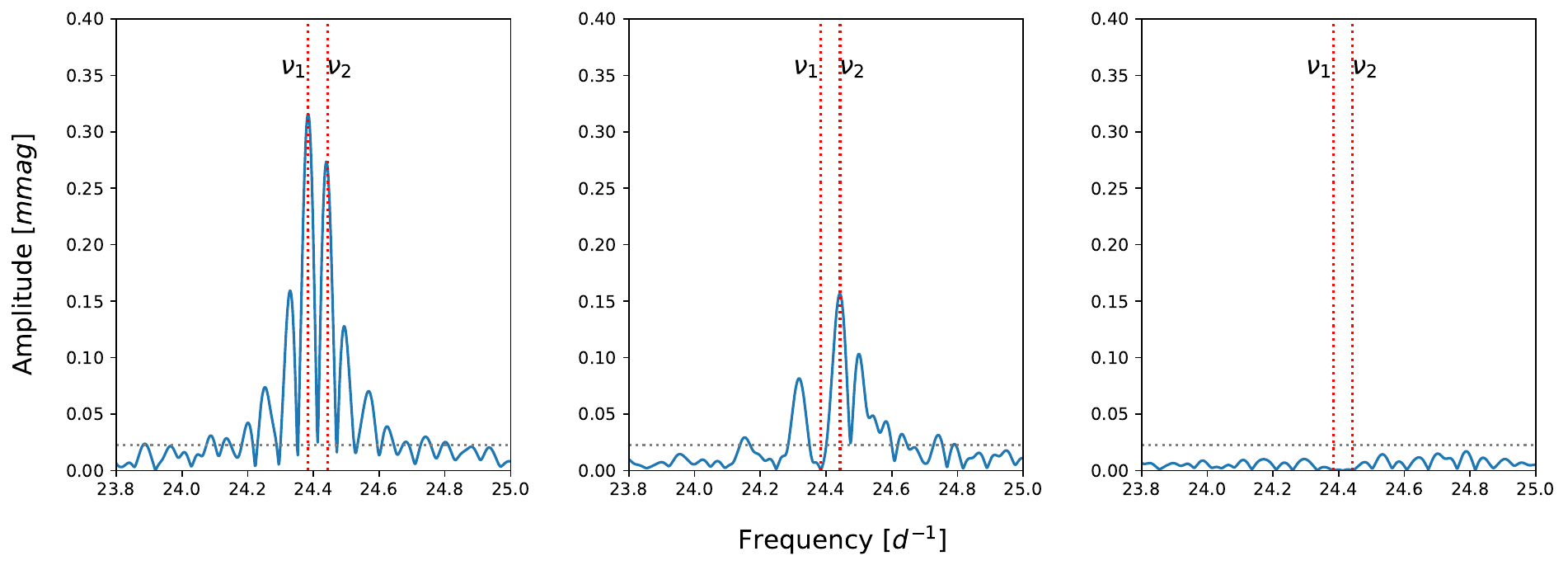}
\bigskip
\begin{minipage}{12cm}

\caption{
\textit{Left panel}: frequency spectra of HD 118660 observed in TESS sector 23 data without performing any pre-whitening, two close frequencies can be seen marked as vertical red dotted lines, \textit{middle panel}: after removing $\nu_{1}$, $\nu_{2}$ can be seen above the confidence level (horizontal black dotted line), \textit{right panel}: after removing both frequencies, no significant signal remained above the confidence level.
Clear evidence of the presence of one close frequency \nub\,=\,24.4420$\pm$0.0008\,d$^{-1}$ is found.
}
\label{fig:prew}
\end{minipage}
\end{figure}

To search for this close frequency, \nua\ was pre-whitened from the original time series. 
In this way, we find evidence for the frequency \nub\,=\,24.4420$\pm$ 0.0007\,d$^{-1}$, which is well resolved from \nua\ according to the Rayleigh criterion (Fig.\,\ref{fig:prew}). After removing two frequencies no other frequency remained above the significance level assuring the fact of the presence of two close frequencies. We can further verify it with a simple two-frequency model.

To confirm whether beating of two close frequencies induces the amplitude modulation, we constructed a two-frequency model on the one-day segmented data by superimposing two sinusoidal signals with frequencies \nua = 24.3837\,d$^{-1}$ and  \nub = 24.4420\,d$^{-1}$, phases  
 $\phi_{1} = 0.5211$ rad and $\phi_{2} = 0.9481$ rad  and amplitudes $A_{1}=0.3169 $ mmag and $A_{2}=0.2309 $ mmag,  calculated with respect to the epoch $T_0$\,=\,2458929.993\,HJD. 
The comparison of the model (red) and observations (blue) are shown in Fig.\,\ref{fig:amp_model}. 
A nearly sinusoidal variation with a frequency of 0.0558\,$\pm$\,0.0147\,d$^{-1}$ (or $17.937\pm 3.753 d$) is clearly visible. 
We, therefore, conclude that the periodic variations observed in the amplitude and phase of \nua\ can be explained by the beating of two close frequencies \nua\,=\,24.3837$\pm 0.0005\,d^{-1}$ and \nub\,=\,24.4420$\pm 0.0008\,d^{-1}$. 

\section{Results}
\label{results}

In this paper, we have presented an analysis of amplitude modulation from the samples observed under N-C survey. 
HD\,118660 is one of them where amplitude modulation has been detected. 
The TESS data set was utilized to investigate the amplitude modulation and monitor the variations in both amplitude and phase at a constant frequency in one day bin. 
It is found that both the amplitude and phase of a single $p$-mode frequency varies with a frequency of 0.0558\,$\pm$\,0.0147\,d$^{-1}$ (see Fig.\,\ref{fig:amp_model}).  
It is seen that the observed modulation frequency and the beating frequency ($|\nu_{1} - \nu_{2}|$) are consistent. 
This confirms that the beating of two close frequencies \nua\ and \nub\ is a valid explanation for the observed periodic amplitude modulation. 
However, the amplitudes and phases of the other frequencies are stable.

\section{Future Prospects}
\label{future}

One of the issues encountered when studying \dsct\,stars is the difficulty of mode identification, i.e., to identify the observed pulsation modes in terms of their radial order $n$, angular degree $l$, and azimuthal order $m$. However, this is one of the basic requirements for an asteroseismic study aiming to probe the internal structure and evolution of the studied stars. 
Our future plan is to identify observed pulsation modes of HD\,118660 with the help of the available TESS photometry and high-resolution spectra, where we are using HERMES data, located at La, Palma, Spain, which will be an application for telescope time within the BINA collaboration. The combined photometric and spectroscopic analysis will be followed in the forthcoming paper. 

Subsequently, we will use stellar evolution codes such as MESA \citep{2015ApJS..220...15P} and pulsation codes such as GYRE \citep{2013MNRAS.435.3406T} for the modelling of HD\,118660 in an attempt to fully understand the multi-periodic pulsational behaviour of this \dsct\,star and add other \dsct\,stars with similar characteristics to our study.

\begin{acknowledgments}
The authors are grateful to the Indian and Belgian funding agencies DST (DST/INT/Belg/P-09/2017) and BELSPO (BL/33/IN12) for granting financial support to organize the third BINA workshop and other BINA activities. MS acknowledges the financial support received from CSIR Fellowship grant 09/948(0006)/2020-EMR-1. 
The data presented in this paper were obtained from the Multimission Archive at the Space Telescope Science Institute (MAST). STScI is operated by the Association of Universities for Research in Astronomy, Inc., under NASA contract NAS5-26555. Support for MAST for non-HST data is provided by the NASA Office of Space Science via grant NAG5-7584 and by other grants and contracts.
\end{acknowledgments}

\begin{furtherinformation}

\begin{authorcontributions}

This work is part of a Indo-Belgium ``BINA''  project where the collective efforts were made by all the co-authors with the relevant contributions.

\end{authorcontributions}

\begin{conflictsofinterest}
The authors declare no conflict of interest.
\end{conflictsofinterest}

\end{furtherinformation}

\bibliographystyle{bullsrsl-en}

\bibliography{extra}

\end{document}